\documentclass[english,aps, achemso,reprint,twocolumn]{revtex4-1}

\usepackage[T1]{fontenc}   
\usepackage{rotating}
\usepackage{booktabs}     
\usepackage[latin9]{inputenc}
\usepackage{graphicx}   
\usepackage{float}
\usepackage{amsmath}
\usepackage{amssymb}
\usepackage{amstext}
\usepackage[colorinlistoftodos]{todonotes}
\usepackage{siunitx}
\usepackage[super]{nth}
\usepackage{commath}
\usepackage{babel} 

\begin{document}

\title{Genetic optimization of training sets for improved machine learning models of molecular properties}

\date{\today}     

\author{Nicholas J. Browning$^{1,2}$}
\author{Raghunathan Ramakrishnan$^{1,3, 4}$}
\author{O. Anatole von Lilienfeld$^{1,4}$}
\email{anatole.vonlilienfeld@unibas.ch}
\author{Ursula R\"othlisberger$^{1,2}$}
\email{ursula.roethlisberger@epfl.ch}

\affiliation{$^1$National Center for Computational Design and Discovery of Novel Materials (MARVEL)}
\affiliation{$^2$Laboratory of Computational Chemistry and Biochemistry, Institute of Chemical Sciences and Engineering. Swiss Federal Institute of Technology (EPFL) CH-1015 Lausanne, Switzerland.}
\affiliation{$^3$Tata Institute of Fundamental Research, Centre for Interdisciplinary Sciences, 21 Brundavan Colony, Narsingi, Hyderabad 500075, India}
\affiliation{$^4$Institute of Physical Chemistry, Department of Chemistry, University of Basel, Klingelbergstrasse 80, CH-4056 Basel, Switzerland}

\begin{abstract}
The training of molecular models of quantum mechanical properties based on statistical machine learning requires large datasets which exemplify the map from chemical structure to molecular property. 
Intelligent \textit{a priori} selection of training examples is often difficult or impossible to achieve as prior knowledge may be sparse or unavailable. 
Ordinarily representative selection of training molecules from such datasets is achieved through random sampling. 
We use genetic algorithms for the optimization of training set composition consisting of tens of thousands of small organic molecules. 
The resulting machine learning models are considerably more accurate with respect to small randomly selected training sets: mean absolute errors for out-of-sample predictions are reduced to $\sim$25\% for enthalpies, free energies, and zero-point vibrational energy,
to $\sim$50\% for heat-capacity, electron-spread, and polarizability, and by more than $\sim$20\% for electronic properties such as frontier orbital eigenvalues or dipole-moments.
We discuss and present optimized training sets consisting of 10 molecular classes for all molecular properties studied. 
We show that these classes can be used to design improved training sets for the generation of machine learning models of the same properties in similar but unrelated molecular sets. 
\end{abstract}

\maketitle   

\section{Introduction}

Experimentally accurate solutions to the time-independent non-relativistic electronic Schr\"odinger equation (SE) $H(\{Z_I,{\bf R}_I\},N_e){\Psi}=\mathit{E}{\Psi}$ for $N_e$ electrons and a collection of atoms involve numerically challenging calculations~\cite{cramer_theories_and_models}. 
This limits routine electronic structure elucidation and accurate high-throughput screening. 
Previous works~\cite{rupp2012fast,ramakrishnan201bigdata,montavon2013machine} have found that the task of repetitiously solving the SE can be mapped onto a computationally efficient, data-driven supervised machine-learning (ML) problem instead. In these models, expectation values of quantum-mechanical (QM) operators are inferred in the subset of chemical space spanned by a set of reference molecular graphs, enabling a speedup of several orders of magnitude~\cite{anatole_fppv} for predicting relevant molecular properties such as enthalpies, polarizabilities and electronic excitations~\cite{montavon2013machine,hansen2015machine,ramakrishnan2015electronic}. Here, QM reference calculations provide training examples $\{(x_\alpha, y_\alpha)\}_{\alpha=1}^N$, where $x_\alpha$ are molecular structures and $y_\alpha$ the expectation values for a chemical property, for interpolation in a 4$N$ dimensional space, aka.~chemical space. 
After training, accurate property predictions for new as of yet unseen molecules can be obtained at the base cost of the underlying ML model, 
provided that the new query molecule lies close to the space spanned by the reference data.

A key issue in the validation of ML models is the selection of appropriate data to use for training.
Here, we will tackle this problem in the context of ML models of quantum chemistry--based estimates of molecular properties. 
Training examples are typically chosen from a uniform random distribution, thus there is no guarantee that the selected data will produce an optimal model. 
In this work we study the effect of optimizing the composition of the set of training examples used for learning, by maximizing the predictive power of the underlying ML model for a given training set size. 

There are numerous alternative strategies to design ML models with reduced relative mean absolute errors (RMAE) of property prediction. 
In previous work k-fold cross-validation has been used~\cite{hansen_assessment_validation} to reduce the predictive error of the ML model by optimizing the hyperparameters $\sigma$ and $\lambda$. 
The molecular representation could also be exchanged without any loss of generality. 
For this study we chose to rely on the sorted Coulomb matrix representation -- a well established and generic representation which meets the crucial uniqueness 
criterion~\cite{FourierDescriptor}. We note that more advantageous descriptors, such as the Bag-of-Bonds (BoB)~\cite{bag_of_bonds}, or Bonds-and-Angles 
Machine Learning (BAML) models~\cite{BAML} could have been used just as well. Finally, instead of focussing on properties directly, $\text{$\Delta$}$-learning~\cite{ramakrishnan201bigdata} can be used to focus on 
the difference between a baseline and  higher level of theory. In this study we include this model for learning the differences between PM7 and B3LYP atomization enthalpies.

This work is based on a quantum chemistry database published in 2014~\cite{ramakrishnan_134k}, 
which contains relaxed geometries and 13 molecular properties computed at the DFT/B3LYP/6-31(2df,p) level of theory for 133'885 small organic molecules of up to 9 heavy atoms, 
extracted from the GDB-17 list of 166B SMILES strings~\cite{GDB_database}, and will herein be referred to as the GDB9 database. 
In this work we use both this and a smaller subset, denoted as the GDB8 database, containing molecules of up to 8 heavy atoms, resulting in 21800 molecules. 
For the GDB8 subset, the following properties are considered for training set composition optimization: enthalpy ${H}$ and free energy ${G}$ of atomization; heat capacity ${C}_v$; isotropic molecular polarizability $\alpha$; electronic spatial extent $\langle R^2 \rangle$; harmonic zero-point vibrational energy $\text{ZPVE}$; energy of the highest occupied $\epsilon_{\text{HOMO}}$ and lowest unoccupied $\text{$\epsilon$}_{\text{LUMO}}$ molecular orbitals; HOMO-LUMO gap $\Delta\epsilon$ and dipole moment $\mu$. 
In addition, the modeling of electronic spectra of GDB8 has been studied in~\cite{ramakrishnan2015electronic}.

\begin{figure*}
\includegraphics[width=0.9\textwidth]{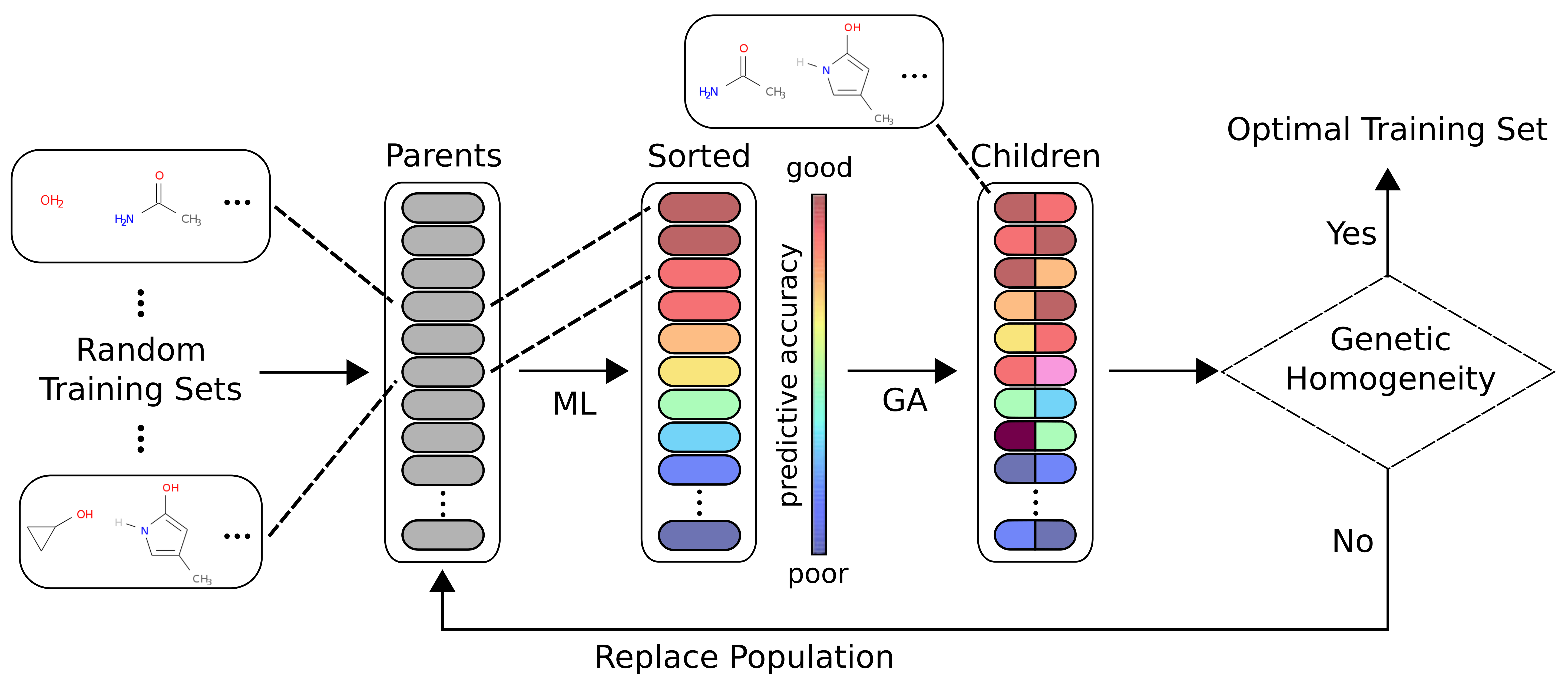}
\caption{
GA procedure to optimize molecular sets for training the ML model.  
A population of trial training sets containing an identical number of molecules are randomly sampled from the molecular database. This initial population of trial training sets serves as the first iteration of the algorithm, and is denoted as the parent population. An ML model is trained for each training set, and the mean out-of-sample prediction error is assigned to the training set as a fitness metric. The population then undergoes fitness-based selection and variation operators to create a child population. A portion of the worst children are replaced by the best parents. Finally the children are labelled as parents, and the algorithm repeats until there exists limited information difference between subsequent iterations.  
}
\label{ga_schematic}
\end{figure*}

\section{Theory}
\subsection{Machine Learning Model}
\label{mlmodel}
Similarly to the Hamiltonian used for electronic structure calculation, molecular information in the ML model has Cartesian coordinate and nuclear charge information inherently encoded through sorted Coulomb matrices~\cite{coulomb_matrix, hansen_assessment_validation}, which are used to represent molecular structures for training and property prediction. The $L^1$ norm $d_{\alpha,\beta} = {|\mathbf{M}^\alpha-\mathbf{M}^\beta|}_{L^1}=\sum_{IJ} |\mathbf{M}^\alpha_{IJ}-\mathbf{M}^\beta_{IJ}|$ then serves as a metric measure of similarity between any two molecules $\alpha$ and $\beta$ in the set of all sorted Coulomb matrices $\{\mathbf{M}\}$. Note that sorted Coulomb matrices encode (except among enantiomers) the external potential of any given molecule uniquely, such that it is invariant with respect to molecular translation, rotation, and atom-indexing. The ML model attempts to construct a non-linear mapping between molecular characteristics and molecular properties. Here, we model the property $P^{\text{ML}}$ of a new out-of-sample molecule \textbf{M} as a linear combination of weighted Laplacian kernel functions $k(\mathbf{M}^{\alpha}, \mathbf{M})$, located on each training instance $\alpha$: 

\begin{equation}
P^{\text{ML}}(\mathbf{M})= \sum_{\alpha=1}^{N}c_\alpha k(\mathbf{M}^\alpha, \mathbf{M})
\end{equation}

\noindent
where $\alpha$ runs over all molecules characterized by sorted Coulomb matricies \textbf{M}$^\alpha$ in the training set of size $N$, and $k(\mathbf{M}^\alpha, \mathbf{M}^\beta) = \exp\left(-d_{\alpha, \beta} / \sigma \right)$ is the Laplacian kernel. For the Coulomb matrix descriptor, the combination of the Laplacian kernel and $L^1$ norm has been shown to yield an optimal combination of low computational cost and good predictive accuracy for models of atomization enthalpies~\cite{hansen_assessment_validation}. The regression coefficient vector $\mathbf{c}$ is obtained by training on $\{\mathbf{M}^\alpha, \mathbf{P}^{\text{ref}}\}$. Note that each molecule $\alpha$ contributes to the property estimate not only according to its distance, but also according to its regression weight $c_\alpha$. The global hyperparameter $\sigma$ corresponds to the kernel-width. The regression coefficients are the solutions to the kernel ridge regression (KRR)~\cite{elem_stat_learning} minimization problem for a given kernel width $\sigma$ and regularization parameter $\lambda$, 

\begin{equation}
\mathcal{L} = {|| \mathbf{P}^\text{ref} - \mathbf{P}^\text{ML} ||}^2 + \lambda \mathbf{c}^\text{T}\mathbf{K}\mathbf{c}
\end{equation}

\noindent
where $\mathcal{L}$ is the Lagrangian to be minimized with respect to the coefficient vector $\mathbf{c}$ and $\mathbf{P}^{\text{ref}}$ is the vector of all reference training data. The solution for the coefficient vector then reads $\mathbf{c} = (\mathbf{K} + \lambda \mathbf{I})^{-1}\mathbf{P}^{\text{ref}}$, with $\mathbf{K}$ being the kernel matrix, and $\mathbf{I}$ the identity matrix. Frequently, 5-fold cross validation is used to estimate optimal values of hyperparameters $\sigma$ and $\lambda$, however, this is computationally expensive to perform during an optimization procedure. Instead, we employ the single-kernel method~\cite{ramakrishnan2015many}, using constant values of $\sigma=1000$ and $\lambda=0$. 

\subsection{Genetic Optimization of Training Sets}
\label{GA_optimization}
Simple systematic enumeration to find training sets which minimize the out-of-sample predictive error of the ML model is a computationally demanding optimization problem. Finding 1000 optimal training molecules to train a machine which can best reproduce the properties of the remaining 20800 molecules in the GDB8 database requires the training of ${21800 \choose 1000}=5.56\times10^{1760}$ machines for a complete search. Clearly, complete optimization of training set composition in such a nearly-infinitely large space is impossible, thus an intelligent search method is required to find near-optimal solutions. Here we employ a genetic algorithm (GA)~\cite{ga_turing,ga_barricelli,ga_fraser}, a biologically inspired meta-heuristic optimization technique, which has proved a successful optimization scheme in highly-dimensional and/or large spaces~\cite{ga_large_space}. The GA optimization process is pictorially summarized in figure~\ref{ga_schematic}. Bounded chemical compound space representations for the ML model, i.e training sets, are represented as a collection of unique molecular pointers, $$\mathbf{X}_{i} = \{x_1, x_2, \dots, x_\alpha, \dots, x_N\}.$$ 
\noindent
\textbf{X}$_i$ is termed a position vector, where each element $x_\alpha$ uniquely maps to each Coulomb matrix $\mathbf{M}_\alpha$, and $N$ is the training set size. Each position vector $\mathbf{X}_i$ is used to train the ML model and therefore facilitates the calculation of the mean absolute error (MAE) of predicting out-of-sample properties. Starting with a population of training sets with unique molecules sampled from a uniform distribution, individuals are stochastically selected using MAE as a fitness criterion. These initial training sets are then successively evolved by applying selection, variation and replacement. Selection determines which training sets should remain in the population to produce the training sets of the following generation. Variation results in new representations of chemical compound space and includes two operations. The first, termed crossover, mixes two training sets uniquely such that two new "child" training sets are produced which contain training molecules from both "parents". The second operator, termed mutation, randomly changes training molecules to introduce new information into the population. Finally, a portion of the worst children are replaced by the best parents to create a new parent population. After a number of iterations, genetic homogeneity is reached within the population and out-of-sample MAEs improve no further. These training sets are termed GA-optimized training sets and are considered to be near-optimal.

\subsection{Training and Validation}
\label{validation}
For the GA optimization of training sets in the GDB8 database, the entire database of 21800 molecules is partitioned into independent training and validation databases, each consisting of 10900 molecules. GA optimization procedures for all properties and training set sizes sample solely from the training database, while the validation set is used for calculating out-of-sample MAEs, used to guide the GA optimization procedure. We also define a third database in which all GDB8 molecules are removed from the GDB9 database, and name it the GDB9-8 database. The GDB9 and GDB9-8 databases have been similarly partitioned into training and validation databases of ~65k and ~55k molecules each, respectively, which are used for rule development detailed in section \ref{GDB9projectionlearning}.

\subsection{Computational Details}
\label{computation_details}
Genetic optimization is performed using a population of 2000 trial training sets for each property and training set size, for a maximum number of 500 iterations. Tournament selection without replacement~\cite{Sastry01modelingtournament} is used with a tournament size of 2. Crossover is performed by partially-matched uniform crossover~\cite{uniform_crossover, ga_spears_dejong, ga_eshelman} with a per-gene crossover probability of 0.5. Mutation is performed by partially-matched polynomial mutation~\cite{Deb96acombined} with a distribution index of 7 and a per-gene mutation probability of $\frac{0.5}{N}$ where $N$ is the training set size. Finally, population replacement is achieved through fitness-proportionate elitism with an elitist ratio of 0.7. These parameters result from rigorous benchmarking, targeting maximal RMAE reduction, and typically result in convergence after $\sim$300 iterations.

\section{Results}

\subsection{GDB8 Learning}
\label{gdb8learning}

\begin{figure}
\includegraphics[width=0.5\textwidth]{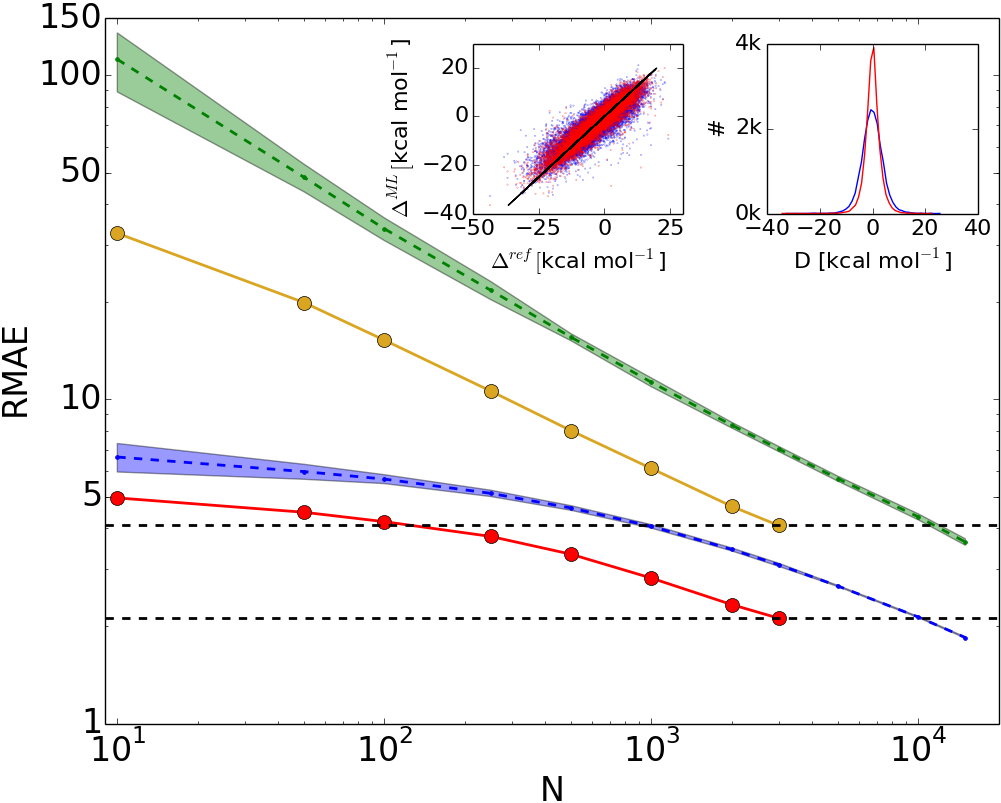}
\caption{Enthalpy of atomization $H$ learning curves using direct-learning $P^{\text{ML}}$ (green), GA-optimized direct-learning $P^{\text{GA}}$ (yellow), B3LYP-PM7 delta-learning $\Delta^{\text{ML}}$ (blue) and GA-optimized delta-learning $\Delta^{\text{GA}}$ (red). Dashed horizontal lines show training set sizes required to reach given accuracies. Left inset: scatter plot of B3LYP-PM7 reference energies of atomization $\Delta^{\text{ref}}$ and predicted values using both a randomly generated $N=1000$ training set $\Delta^{\text{ML}}$ (blue), and the GA-optimized counterpart $\Delta^{\text{GA}}$ (red). Right inset: typical error distribution over B3LYP-PM7 reference atomization enthalpies for aforementioned random (blue) and GA-optimized (red) $N=1000$ training sets. }
\label{rmae}
\end{figure}

\begin{table*}
\setlength\extrarowheight{2pt}
\caption{Randomized and GA-optimized out-of-sample RMAEs for all properties. All target chemical accuracies are 1 kcal mol$^{-1}$, except for ZPVE, dipole moment and isotropic polarizability, which have target accuracies of 10cm$^{-1}$, 0.1$D$ and 0.1$a_0^3$ respectively. GA-optimized RMAEs are denoted by $P^{\text{GA}}$ while randomly generated training set RMAEs are denoted as $P^{\text{ML}}$. Final row corresponds to out-of-sample RMAEs for enthalpy of atomization $H$ using $\Delta$-learning (aka $\Delta H_{\rm PM7}^{\rm B3LYP}$), and bracketed the GA optimized counterpart, referred to as $\Delta^{\text{ML}}$ and $\Delta^{\text{GA}}$ respectively.}
\begin{tabular}{l rrrrrrrrrrrrrrrrrrrrr  }
\toprule
     \textit{P}$^{\text{ML}}$ (\textit{P}$^{\text{GA}}$) ~~~~& \multicolumn{21}{l}{\textit{N}} \\
     \cline{2-22}
     & \multicolumn{3}{l}{ 10} & \multicolumn{3}{l}{50}  & \multicolumn{3}{l}{100}  & \multicolumn{3}{l}{500}  & \multicolumn{3}{l}{1k}  & \multicolumn{3}{l}{2k} & \multicolumn{3}{l}{3k}  \\
   \midrule
     \textit{H} & 113.0 & (31.6) & & 48.0 & (18.3) & & 33.3 & (14.3) & & 14.8 & (7.5) & & 10.2 & (5.8) & & 6.8 & (4.5) & & 5.1 & (3.9) & \\
   \hline
     \textit{G} & 101.8 & (28.8) & & 44.0 &(17.7) & & 31.4 & (14.1) &  & 14.3 & (7.5) & & 9.9 & (5.6) & & 6.7 & (4.3) & & 5.0 & (3.9) & \\
   \hline  
     \textit{C}$_v$ & 27.3 & (14.5) & & 18.2 & (9.4) &  & 14.6 & (7.8)  & & 7.4 & (4.0) & & 5.2 & (2.9) & & 3.4 & (2.3) & & 2.5 & (2.0) & \\
    \hline    
     ZPVE  & 353.2 & (83.2) & & 150.4 & (38.4) & & 97.9 & (28.0) & & 31.5 & (14.0) & &  20.9 & (10.5) & & 13.9 & (7.0) & & 10.5 & (3.5) & \\
    \hline  
    <$\text{R}^2$> & 168.5 & (92.2) & & 117.0 & (44.2) & & 85.6 & (33.2) & & 35.7 & (19.1) & & 25.7 & (15.5) & & 18.3 & (12.7) & & 14.5 & (11.6) & \\
    \hline
      $\mu$ & 11.3 & (8.5) & & 10.3 & (7.7) & & 9.9 & (7.4) & & 8.4 & (6.3) & & 7.5 & (5.7) & & 6.2 & (5.1) &  & 5.2 & (4.7) & \\
    \hline    
      $\alpha$ & 40.8 & (16.3) & & 23.1 & (12.0) & & 18.5 & (10.8) & & 11.8 & (7.8) & & 9.6 & (6.5) & & 7.2 & (5.4) & & 5.8 & (4.9) & \\
    \hline  
     $\epsilon_\text{HOMO}$ & 13.0 & (9.0) & & 11.2 & (8.1) & & 10.4 & (7.3) & & 7.7 & (5.2) & & 6.3 & (4.5) & & 4.9 & (3.8) & & 4.0 & (3.5) & \\
    \hline  
     $\epsilon_\text{LUMO}$ & 22.3 & (15.8) & & 18.8 & (12.7) & & 17.0 & (11.1) & & 11.9 & (8.0) & & 9.7 & (6.7) & & 7.4 & (5.6) & & 5.9 & (5.0) & \\
    \hline  
    $\text{gap}$ & 24.0 & (17.8) & & 20.8 & (15.0) & & 19.5 & (13.5) & & 14.3 & (9.8) & & 11.8 & (8.1) & & 9.0 & (6.8) & & 7.3 & (6.2) & \\
   \hline
    $\Delta^{\text{ML}}(\Delta^{\text{GA}}) H$ & 6.6 & (5.0) & & 6.0 & (4.4) & & 5.7 & (4.1) & & 4.6 & (3.2) & & 4.1 & (2.6) & & 3.4 & (2.1) & & 3.1 & (1.9) & \\
   \bottomrule
\end{tabular}
\label{all_data}
\end{table*}

All reported relative mean absolute errors (RMAEs) refer to out-of-sample prediction of $10900$ molecules from a machine trained on $N$ in-sample molecules. 
Target accuracies (for which RMAE = 1) for thermochemical properties and orbital energies are 1 kcal mol$^{-1}$. 
For ZPVE a target accuracy of 10cm$^{-1}$ was selected, comparable to the average accuracy of coupled cluster methods 
with converged basis sets~\cite{tew_ccsd_basis_lim} for predicting harmonic vibrational frequencies of small molecules. 
For isotropic polarizability and norm of dipole moment, 
target accuracies of $0.1 a_0^3$ and $0.1$D were used. 
Again, these values are within the predictive uncertainty using CC level of theory~\cite{hickley_dm_ip_accuracy}.

Figure~\ref{rmae} displays learning curves of randomly generated and GA-optimized training sets for out-of-sample predictions of B3LYP level enthalpies of atomization, $H$, using both direct learning as well as the $\Delta H_{\rm PM7}^{\rm B3LYP}$-ML model~\cite{ramakrishnan201bigdata}. 
Due to KRR based ML errors decaying as inverse powers of training set size, we present RMAEs as a function of training set size on a log-log scale. 
Upon GA optimization we note a substantial lowering of the learning curves for both properties and all training set sizes.
The combination of GA optimization with $\Delta$-ML model is the most promising: 
For machines trained on 3k molecules the GA optimization reduces the RMAE from 3.1 to 1.9. 
Corresponding scatterplots and error distribution plots are on display as insets in figure~\ref{rmae}.
They also indicate that the GA-optimized $\Delta$-ML models approach the ideal model much faster than randomly sampled training sets.
It is particularly encouraging to note that the relative gain in predictive power, obtained by using GA, 
appears to converge towards a constant rather than to vanish.

\begin{figure*}
\includegraphics[width=\textwidth]{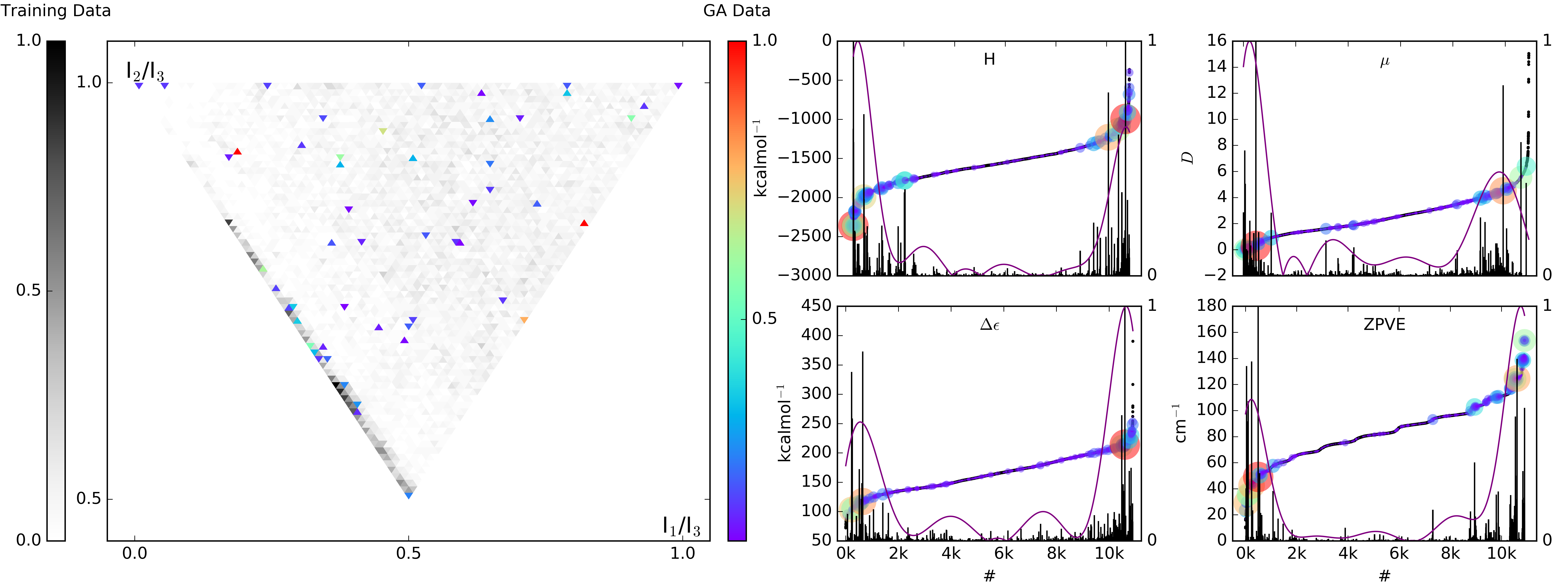}
\caption{Left: Principal moment of inertia density plots for the training database (grayscale) and 1000 GA-optimized N=10 training sets (colored) for enthalpy of atomization, H. Color corresponds to selection probability upon GA optimization. Right: training data cumulative density plots for the properties: H, $\mu$, $\Delta\epsilon$, and ZPVE. Normalised selection probabilities inferred from the ensemble of GA optimisations for $N$=10 shown in black, with a corresponding \nth{8} order polynomial fit shown in purple. Circle sizes and colors also correspond to selection probabilities.}
\label{selection_pressure}
\end{figure*}

RMAEs for ML models with and without using GA 
are summarized in table \ref{all_data} for various training set sizes and all the aforementioned properties.
GA optimization of training set composition systematically improves RMAEs for all properties and training set sizes. 
Some properties experience greater improvement than others. 
The reduction in RMAE is most prevalent for smaller training set sizes, particularly of size $N$=10. 
More specifically, properties related to chemical bonding, such as enthalpies and free energies of atomization as well as ZPVE improve by $\sim$75\%.
Other extensive properties, such as heat capacity or polarizability, improve by $\sim$50\%.
In contrast, intensive electronic properties experience much less improvement.
Overall, the smallest RMAE reduction is found for the norm of the dipole moment.
While percentage wise, seemingly small, the error reduction for $\Delta H_{\rm PM7}^{\rm B3LYP}$ is still very relevant due to its outstanding accuracy in absolute terms. 
This finding could possibly be also due to the comparatively more demanding target accuracies, 
the standard deviation of energetic properties in the data set is 
orders of magnitude larger than the standard deviation of electronic properties, such as the dipole moment. 
See figure~\ref{selection_pressure} for comparison.
   
Overall, these results clearly amount to numerical evidence that the choice of how to select training set members 
can have a dramatic effect on the predictive power of the resulting ML model. 
It follows that substantially fewer training examples would be needed for the generation of ML models which reach the same accuracy 
as ML models trained on a much larger training set sampled at random. 
In order for this insight to be useful, however, one would also require access to the solution of the selection optimization
process in an \textit{a priori} fashion, and not {\em a posteriori} as it is the case in this study. 

\begin{figure*}
\includegraphics[width=\textwidth]{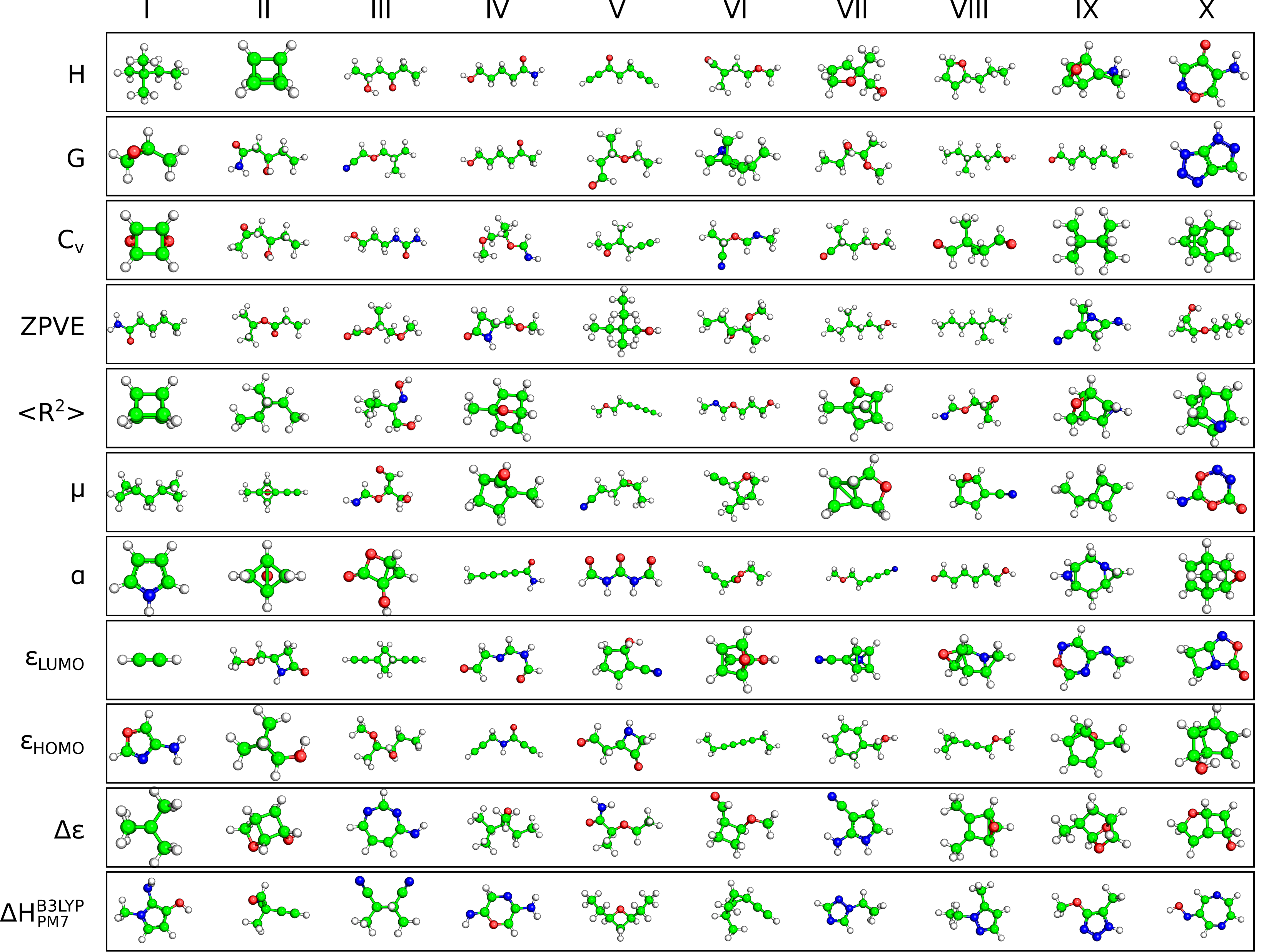}
\caption{Molecular classes I-X for all properties listed in table \ref{all_data}. These classes correspond to the N=10 column, where each molecule and its respective training set is collectively presumed to optimally represent the entire space of validation molecules. Each training set is sorted by its index in the GDB8 database, thus chemical weight approximately increases from left to right.}
\label{10_trainingsets}
\end{figure*}

\subsection{Analysis of GA-Optimized GDB8 Training Sets}
\label{gdb8n10training}

The additional GA optimization layer can be seen as providing the underlying machine learning model with the capacity to intelligently select its own data for optimal out-of-sample prediction. 
This in turn naturally means that the machine, through the kernel, $L^1$ distance metric and CM representation, is finding optimal maps from molecular structure and composition to property with respect to the diverse chemical space of the validation database.  
As such, it is interesting to inspect the outcome of this optimization.
While this is rapidly overwhelming for larger training sets, for $N$=10 training sets the most essential chemistry encoded can easily be grasped.
Figure~\ref{10_trainingsets} shows the optimized sets for all properties. 

It is important to keep in mind that the molecules shown likely do not represent the global optimum, rather they represent one of many near-optimal combinations. We have obtained them by first resampling unique ensembles of $N$=10 training sets many times, with a biased probability proportional to the $L^1$ distance from each molecule in the best GA-optimized training set for each property. All training sets sampled in this way produced significantly better RMAEs comparatively to random sampling.  Thus, each GA-optimal molecule should be interpreted as being representative for very similar compounds which could be selected in optimized training sets, and are reported in figure~\ref{10_trainingsets}.
We therefore find it appropriate to relabel each index in the GA-optimal training sets as molecular classes I-X. 
Note though that, due to lack of an iterative optimisation procedure, ML models based on such biased training sets composed of these similar molecules yield slightly worse performance than ML models resulting from the underlying GA-optimized training sets. 

From inspection it is obvious that the chemistry, represented by these classes, is very rich, including linear, planar, cage like, branched, and even strained structures. 
Variety in chemical composition is maintained through the occasional replacement of carbon units by functional groups containing oxygen or nitrogen.
Furthermore, all hybridization states, $sp^3$, $sp^2$, and $sp$ are represented.
It can also be seen that many properties share very similar molecular classes, with some molecules being shared across optimal training sets for different properties. 
For example classes VII and VIII for ZPVE and $G$, or classes VIII and IV for $\alpha$ and $G$, respectively.
We note that for energetic properties in particular, very stable and unstrained saturated molecules are selected as well as rather
exotic unsaturated and strained systems, suggesting a bias towards the extremes of the distribution.  Additionally, these exotic molecules tend to appear more so for more complex properties, such as HOMO-LUMO gap or $\Delta H_{\rm PM7}^{\rm B3LYP}$.
However, we do not find it obvious to further rationalize the specific selection of chemistries.

\subsection{GDB8 Selection Pressures}
\label{gdb8selection}
While the visual interpretation of the chemical classes discussed previously is not obvious, 
we can use the statistics from the GA optimization runs to systematically identify the effect of the bias. 
More specifically, to understand selection pressures we analyzed the regions of chemical space which produce optimal training sets for a given training set size. 
To this end, we have run 1000 identical GA optimizations for training sets of size $N$=10 to obtain sufficient statistics to infer selection probabilities. 
In figure~\ref{selection_pressure} we plot these results in terms of the principal moment of inertia density of molecular structures (left) for enthalpies of atomization, $H$, 
as well as in terms of the cumulative density plots (right) for properties $H$, $\mu$, $\Delta\epsilon$ and ZPVE. Additionally on each cumulative density plot, we overlay the inferred selection probabilities of all molecules within the training database. 
It can be seen from the principal moment of inertia density plot, that while the training database disproportionately contains linear and oblate/prolate structures, 
GA optimization does not necessarily drive training set composition towards regions of high density within the GDB8 database. 
Indeed, many of the molecules which are frequently selected during training set composition optimization are from medium to low density regions and appear to 
rather homogeneously cover the entire plane. 

\begin{figure}
\includegraphics[width=0.5\textwidth]{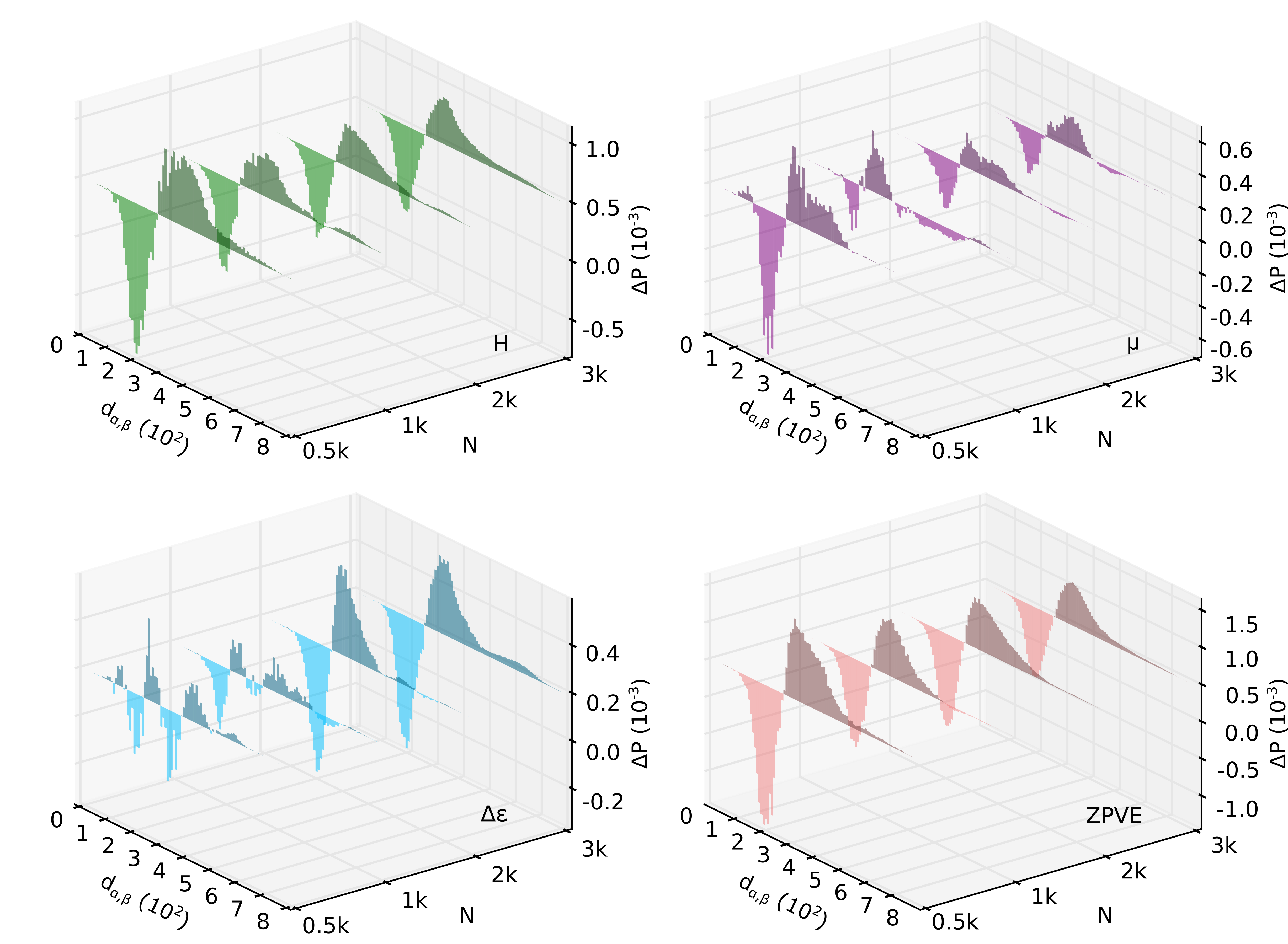}
\caption{GA-optimization tends to increase molecular distances: 
Normalised $L^1$ distance distribution differences (GA-optimized training set distribution minus randomly selected counterpart) for properties $H$ (top left, green), $\mu$ (top right, purple), $\Delta\epsilon$ (bottom left, cyan) and ZPVE (bottom right, orange) as a function of training set size $N$. }
\label{distance_distrib}
\end{figure}

In contrast, from the cumulative property density plots it can be seen that GA optimization preferentially 
selects regions of low property density (tails), while regions of high density (linear) are less likely to be sampled. 
We have investigated if introducing a deliberate bias into the training set sample, 
through importance sampling of the property distribution alone (shown in figure~\ref{selection_pressure}) 
to specifically over-represent these regions, affords ML models with increased predictive power. 
Unfortunately, the introduction of such bias does not necessarily yield any improvement over random sampling (and indeed can even be worse). 
Thus, while there are systematic trends upon GA optimization, there do not seem to be simple biasing rules purely based on properties,
and the effective change in sampling chemical space has to be taken into account.
We illustrate the effect of the GA-optimization on training set distribution in figure~\ref{distance_distrib}: 
We plot the difference in $L^1$ distance distribution between an average of 1000 randomly generated training sets of various sizes and their GA-optimized counterparts for the same properties. 
It can be seen that there is a small yet systematic GA-induced outward shift, i.e.~to predominantly sample  training molecules which are further apart from each other across chemical space. 
As with the reduction in RMAE upon GA-optimization these changes are not identical across properties, and appear to slowly converge to a constant $\Delta P$ per property. In view of this, we note that we have additionally tried to optimize training sets through maximal $L^1$ spread and thus generate training sets which are more chemically diverse, however, this yielded no improvement over random sampling.

\subsection{Design and Application of $N$=10 Training Sets for GDB9-8 ML Models}
\label{GDB9projectionlearning}

\begin{figure}
\includegraphics[width=0.5\textwidth]{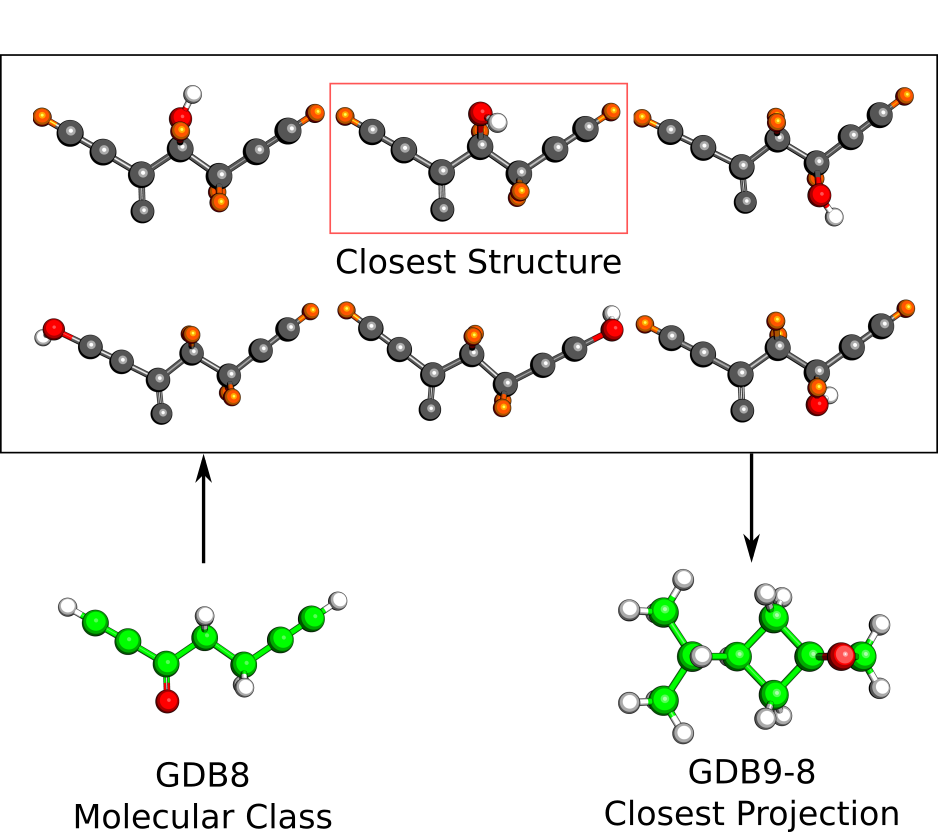}
\caption{
Exemplification of applying the projection rule GDB8 $\rightarrow$ GDB9-8 to the molecular class $H$-V (lower left, see figure ~\ref{10_trainingsets}) for -OH substitution. 
The box contains the pseudo-GDB9-8 molecules obtained by substitution of each alkyl hydrogen (shown in orange). 
For each of these pseudo molecules the entire GDB9-8 database is queried and the closest molecule is selected to represent the projection of this molecular class into the GDB9-8 database
(lower right). 
}
\label{projection}
\end{figure}

In order to investigate the transferability of the above insights gained upon the molecular classes 
we have studied the effect of designing an artificially constructed training set, derived from the classes discovered through the use of GA on the GDB8 database.
More specifically, we define "projection" rules to sample the larger and independent GDB9-8 database for which no GA has ever been performed. 
For each of the GDB8 molecules used to train the ML model of $H$, on display in figure~\ref{10_trainingsets}, 
we individually replace each hydrogen with the heavy substituents -OH, NH$_2$ and CH$_3$.
Figure ~\ref{projection} illustrates the procedure for the projection of GDB8 molecule $H$-V in figure~\ref{10_trainingsets} to GDB9-8. 
This procedure leads to hundreds of pseudo-GDB9-8 molecules, all having the same number of heavy elements as the other molecules present in GDB9-8, containing a total of $\sim$100k molecules. 
For each of these new, pseudo-GDB9-8 structures, we search the GDB9-8 training database 
(discussed in section \ref{validation}) 
for the closest set of molecules which represent each of the ten classes.
With the resulting molecules we trained new ML models for $H$ which are applicable to GDB9-8. 

While there are many pseudo-GDB9-8 structures found during the search which do exist (with some deviation in $L^1$ due to small geometrical differences), many of them do not have real GDB9-8 counterparts and in these cases a closest analogue is selected. When creating training sets for use in the GDB9-8 database, we do not weight these unfavourably and instead simply generate them with molecules uniformly sampled from each projected molecular class. On average, the resulting ML model affords a reduction of RMAE for out-of-sample predictions by more than 20\% (decrease from 116.2 to 90.4) with respect to randomly generated training sets. Furthermore, upon use of the projected molecular classes, the error of the most extreme outlier in GDB9-8 is reduced from 275.4 to 116.1, while the error of the best training set reduces from 75.3 to 61.4.
One should keep in mind that this still represents a substantial reduction: Due to the exponential scaling of chemical space with molecular size, GDB9-8 is roughly an order of magnitude larger than GDB8. 
As such, one should not expect that a simple projection from the 10 best molecules in GDB8 towards GDB9-8 results in the optimal set of 10 molecules out of GDB9-8,
but rather in substantial sub-optimality.

\section{Conclusions}
Here, genetic algorithms (GA) have been employed to optimize the composition of molecular training sets that can be used for the training of machine learning models of molecular properties. 
Application of GA to a training set of a given size improves the performance of the ML model substantially, when compared to a model trained on randomly selected molecules coming from the GDB subspace of organic molecules that follows a distribution that is comprehensive with respect to chemical intuition. 
Conversely, for achieving the same accuracy, dramatically less training examples are necessary---provided that the user has the possibility to bias the selection of training examples prior to training. 
Ensembles of GA optimization procedures have shown that, while there is evidence of systematic bias towards low density regions of property distributions, attempting to construct a beneficially biased training instance distribution for improved ML models is not obvious.
However, the design of improved training sets containing structures identified through the use of simple molecular projection rules applied to GA-optimized training sets 
seems to be possible. We have exemplified this for $\sim$100k molecules in GDB9-8 using constructed training set molecules obtained by GA for $\sim$20k molecules in GDB8.

It is worth noting that all optimized training sets reported are unlikely to correspond to global optima, but instead are rather near-optimal. 
The collection of training molecules producing such optimized RMAEs is heavily dependent on the choice of GA parameters, but in particular upon population size, i.e the extent of training set sampling. Conversely the speed of optimization is directly linked to population and training set sizes, thus there is a trade-off between RMAE reduction and computation time. 
Nevertheless this error can be further reduced using larger population sizes, or more training data. Tightening GA convergence criteria may also yield some improvement. 
Furthermore, the distribution of optimal training instances depends strongly on the chosen fitting function, in our case the specific combination of Slater-type kernel function, Manhattan distance, and Coulomb-matrix. 
Modification of the model (through representation/regressor/hyperparameter-selection) will likely influence the details of the selection 
pressure of particular molecules for a given training set size and property.
Any other ML model is likely to lead to another optimal distribution. 
However, the overall insight that substantial model improvements can be obtained through GA optimization should be general 
since the selection bias present in training sets is independent of model details. 
The exact nature of the relationship between ML model specifics and selection bias remains to be elucidated in future work. 

Finally, we note that while we have investigated how to remove the bias in a training set of a certain size for a given ML model, 
the inherent bias in the entire data set has not been explored---it is an open question, however, if that is even possible, 
due to the practically infinite number of possible molecular compositions and geometries.
Future work will show if our conclusions also apply to ML models trained on other subsets of chemical space,
e.g.~corresponding to ZINC-molecules or the ``representative set of molecules'' ~\cite{baratan_chem_space}.

\section{Acknowledgments}
This research was supported by the NCCR MARVEL, funded by the Swiss National Science Foundation.
OAvL acknowledges funding from the Swiss National Science Foundation (No.~PP00P2\_138932).
Some calculations were performed at sciCORE (http://scicore.unibas.ch/) Scientific Computing
Core Facility at the University of Basel and at the Swiss National Supercomputing Centre (CSCS, http://www.cscs.ch/).

\bibliographystyle{achemso}
\bibliography{paper}

\end{document}